\documentclass[namedreferences]{solarphysics}
\usepackage[hyperref,optionalrh,solaromanenum]{spr-sola-addons} 
\usepackage{graphicx}                    
\usepackage{color}                       
\usepackage{breakurl}                         

\newcommand{\aap}{    {\it Astron. Astrophys.}}
\newcommand{\aaps}{   {\it Astron. Astrophys. Suppl.}}

\newcommand{\apj}{    {\it Astrophys. J.}}
\newcommand{\apjl}{   {\it Astrophys. J. Lett.}}

\newcommand{\pasp}{   {\it Pub. Astron. Soc. Pac.}}

\newcommand{\solphys}{{\it Solar Phys.}}
 
\newcommand{\ssr}{    {\it Space Sci. Rev.}}

\begin{document}

\begin{article}

\begin{opening}

\title{Polarization of the Corona Observed During the 2017 and 2019 Total Solar Eclipses}

%
\author[addressref={aff1,aff2},corref,email={yoichiro.hanaoka@nao.ac.jp}]{\inits{Y.H.}\fnm{Yoichiro}~\lnm{Hanaoka}\orcid{0000-0003-3964-1481}}
\author[addressref={aff2,aff3}]{\inits{Y.S.}\fnm{Yoshiaki}~\lnm{Sakai}}
\author[addressref=aff4]{\inits{K.T.}\fnm{Koichi}~\lnm{Takahashi}}

%
\runningauthor{Y. Hanaoka et al.}
\runningtitle{Polarization of the Corona}

\address[id=aff1]{National Astronomical Observatory of Japan, 2-21-1 Osawa, Mitaka, Tokyo, 181-8588, Japan}
\address[id=aff2]{Solar Eclipse Digital Imaging and Processing Network, Tsukuba, Ibaraki, 305-0813, Japan}
\address[id=aff3]{Chiba Prefectural Ichihara High School, Ichihara, Chiba, 290-0225, Japan}
\address[id=aff4]{NPO Kwasan Astro Network, 17-1 Ohmine-cho Kitakazan, Yamashina-ku, Kyoto 607-8471, Japan}

\begin{abstract}
We carried out polarimetric observations of the white-light corona during the total solar eclipses that occurred on 2017 August 21 and 2019 July 2, and successfully obtained data at two different sites for both eclipses. After eliminating the sky background, we derived the brightness, polarization brightness, and degree of the polarization of the K+F corona from just above the limb to approximately 4 $R_\odot$. Furthermore, we isolated the K- and F-corona with plausible degree of polarization of the K-corona. The field of view covering up to approximately 4 $R_\odot$ enabled us to compare the derived brightness and polarization with a wide range of other observations. The results of the comparison show significant scatter; while some of the observations present very good coincidence with our results, the other ones exhibit systematic discrepancy. 
\end{abstract}

%

\end{opening}

%
\section{Introduction} \label{sec:intro}

Total solar eclipses provide us very low sky-background in visible and near infrared wavelengths down to just above the solar limb, which cannot be achieved by other means. Therefore, they are scientifically valuable even today. That is why many attempts to observe total solar eclipses in order to obtain scientific data of the solar corona have been made. The eclipse of 2017 August 21, of which the umbra swept across the North America over about 4000 km, was a good example; a variety of observations were made during the 2017 eclipse \citep[see][references therein]{2018A&G....59d4.19P}. In addition, accumulations of modern observations of eclipses enabled studies beyond snapshot-type analyses of individual eclipses \citep[e.g.,][]{2020ApJ...895..123B, 2021ApJ...911L...4H}.

The white-light corona of the Sun has been one of the most important targets for the total solar eclipse observations. Meanwhile, in recent decades, ground-based and spaceborne coronagraphs, such as  COronal Solar Magnetism Observatory (COSMO) K-coronagraph \citep[K-Cor;][]{2013ApJ...774...85H} of the Mauna Loa Solar Observatory (MLSO) on the ground and the Large Angle Spectrometric Coronagraph \citep[LASCO;][]{1995SoPh..162..357B} of the Solar and Heliospheric Observatory (SOHO) and the Sun Earth Connection Coronal and Heliospheric Investigation (SECCHI)-COR1/2 \citep{2008SSRv..136...67H} of the Solar Terrestrial Relations Observatory (STEREO) in space, have also observed it almost regularly. The hot plasma in the corona can be observed as the K-corona, which is produced by the Thomson scattering of the light emitted from the solar disk caused by the free electrons. Although it is the primary component of the white-light corona, the F-corona from interplanetary dust is also contained in the white-light corona. Furthermore, in ground-based observations, the sky background overlaps images of the corona.

Linear polarization enables the separation of observed brightness into the K- and F-corona \citep{1950BAN....11..135V}, as well as the sky background. The K-corona exhibits strong tangential polarization, while the polarization of the F-corona is believed to be negligible. The sky exhibits completely different distribution of polarization from that of the K-corona.
Therefore, polarimetry has long been performed during total eclipses, and today, the coronagraphs mentioned above, which routinely take white-light images of the corona, also take the polarization information \citep[see][for a summary of the polarimetry of the white-light corona]{2020SoPh..295...89L}. Particularly, for the ground-based white-light coronagraphs, polarization is the primary target of the observation, because it is difficult to obtain ordinary white-light images of the corona because of the brightness of the sky background.

The brightness distribution of the K-corona represents the structure and amount of coronal plasma regardless of its temperature, unlike that of extreme ultraviolet or X-ray emissions. The electron density distribution in the corona can be estimated from the polarization data of the K-corona using a model discussed by \cite{1950BAN....11..135V}. In the recent ``digital era'', sophisticated electron-density estimations have been carried out in, e.g., \cite{2001ApJ...548.1081H}, \cite{2002A&A...393..295Q}, and \cite{2012SoPh..277..267S} using eclipse and coronagraph data.

Therefore, the observation of the K-corona is expected to contribute to the research on a coronal plasma-producing mechanism. In addition, it contributes to the study on the coronal variation according to the solar activity cycle. In fact, the long-term variability of the white-light corona has been studied using eclipse data \citep[e.g.,][]{1989A&AS...77...45L, 2014CoSka..44..119R}. Recently, it was also studied using LASCO's observations \citep{2015SoPh..290.2117B, 2020SoPh..295...89L, 2020SoPh..295...20B}. 

Although the above mentioned coronagraphs regularly obtain white-light data, the coronagraph observations are limited in obtaining the complete picture of the corona. The currently operated spaceborne coronagraphs solely observe outer corona owing to the large radii of their occulting disks (2.2 $R_\odot$ and 1.4 $R_\odot$ for LASCO C2 and SECCHI COR1, respectively). In contrast, K-Cor can only observe the inner corona owing to the bright sky background.

However, both the brightness and polarization of the white-light corona, from just above the solar limb to the elongation of several solar radii, can be measured during total solar eclipses, under a very low sky background. Furthermore, the accuracy of polarimetry during eclipses has been significantly improved \citep{1999SoPh..190..185L, 2012SoPh..277..267S, 2012SPIE.8450E..40C, 2020PASP..132b4202V, 2019SoPh..294..166J}. Therefore, the observations of total solar eclipses remain valuable. The wide coverage of eclipse observations can bridge the results from spaceborne and ground-based coronagraphs, and intercomparison is possible between the results obtained from eclipse and coronagraph observations. Such intercomparison is indispensable in obtaining the accurately-calibrated complete picture of the white-light corona.

During two total solar eclipses on 2017 August 21 and 2019 July 2, we conducted multi-site observations of the white-light corona by taking advantage of professional-amateur collaborations. We performed similar observations during former eclipses, and obtained scientific results \citep{2012SoPh..279...75H, 2014SoPh..289.2587H, 2018ApJ...860..142H}. During the 2017 and 2019 eclipses, we also carried out polarimetry on the corona. In this paper, we present the brightness and polarization of the white-light corona measured during the two eclipses, and discuss the comparison of our measurements with other eclipse data and those taken with coronagraphs. Observations and data processing are presented in Section \ref{sec:obsdataproc}, and the quantitative results obtained from the analysis are described in Section \ref{sec:results}. Finally we summarize and discuss the results in Section \ref{sec:summary}.

\section{Observations and Data Processing} \label{sec:obsdataproc}

\subsection{Instruments and Observations} \label{subsec:instobs}

\begin{table}
\caption{Observation sites, eclipse circumstances, and instrumental parameters.}\label{tbl:table1}
\begin{tabular}{lllll}   
\hline
Date & \multicolumn{2}{c}{2017 Aug. 21} & \multicolumn{2}{c}{2019 Jul. 2} \\
\hline
\multicolumn{5}{l}{Observation site} \\
Name & Salem & Madras & CTIO\tabnote{Cerro Tololo Inter-American Observatory} & Elqui \\
 & Oregon, USA & Oregon, USA & Coquimbo, Chile & Coquimbo, Chile \\
Latitute & 44.94N & 44.67N & 30.17S & 29.15S \\
Longitude & 123.03W & 121.14W & 70.81W & 70.89W \\
\hline
\multicolumn{5}{l}{Eclipse circumstances} \\
Maximum & 17:18:18 UT & 17:20:37 UT & 20:39:36 UT & 20:40:15 UT \\
Duration & 1m54s & 2m03s & 2m04s & 1m39s \\
Altitude & 39.9$^\circ$ & 41.6$^\circ$ & 13.0$^\circ$ & 13.5$^\circ$ \\
Z.A.(celestial)\tabnote{Zenith angle relative to the celestial north} & -40.1$^\circ$ & -39.3$^\circ$ & 131.2$^\circ$ & 130.3$^\circ$ \\
Z.A.(solar)\tabnote{Zenith angle relative to the solar north} & -58.3$^\circ$ & -57.5$^\circ$ & 133.2$^\circ$ & 132.3$^\circ$ \\
\hline
\multicolumn{5}{l}{Instrumental parameters} \\
Camera & NikonD810a & CanonEOS6D & NikonD810a & CanonEOS6D \\
Number of pixels & $7380\times 4928$ & $5496\times 3670$ & $7380\times 4928$ & $5496\times 3670$ \\
Pixel scale & $1''.69$ & $2''.27$ & $1''.69$ & $2''.27$ \\
\#Set\tabnote{Number of obtained (valid) polarization datasets. Part of the data at Salem were omitted because an airplane passed through the field of view.} & 3+1/3 & 4 & 4 & 3 \\
\hline
\end{tabular}
\end{table}

We conducted polarization measurements of white-light corona at four sites during the 2017 eclipse and at three sites during the 2019 eclipse. We successfully obtained data under good weather and instrument conditions at two of the sites for each eclipse. The sites, where the data used for the analysis were obtained, are presented in Table \ref{tbl:table1}, including the eclipse circumstances at these sites and instrumental parameters. 

We prepared several, mostly identical instruments for the multi-site polarimetry observations, which include a small telescope, filter wheel, and digital single-lens-reflex (DSLR) camera. The telescope adopted was Takahashi FS-60Q, with an objective lens diameter and focal length of 60 and 600 mm, respectively. The field of view spans an area of $3^\circ .4 \times 2^\circ .3$ ($13\times 9$ $R_\odot$, covering the elongation of $>$ 4 $R_\odot$). A DSLR camera has a detector with Bayer filters for color imaging, and can be used to obtain broadband images of the white-light corona in red (R), green (G), and blue (B) color-channels. Their central wavelengths are approximately 620, 530, and 470 nm, respectively, for Nikon D810a, and 600, 530, and 470 nm, respectively, for Canon EOS6D. As described in the following sections, the polarization information, as well as multi-color data, enabled us to separate the observed brightness and polarization into K-corona, F-corona, and sky background.

The filter wheel (fabricated by Koheisha, Kawagoe, Japan) was attached between the telescope tube and the camera. In the filter wheel, we installed three linear polarizers, which have transmission axes of $0^\circ$, $60^\circ$, and $-60^\circ$, respectively, relative to the horizontal axis of the camera frame. In addition, a neutral density filter, which has approximately the same thickness and transmission as the polarizers, was also installed in the filter wheel to obtain ordinary white-light coronal images. In this conventional modulator, which is the same type as that of LASCO and SECCHI, the polarization state is altered with the rotation of the filter wheel. In our eclipse observations, we took five images with different exposure times, 1, 1/4, 1/16, 1/64, 1/256, and 1/1024 s, for each filter/polarizer. It takes approximately 30 s to obtain a set of data with the neutral density filter and three polarizers. During the totality period, 3--4 sets of the polarization data were taken at each site, as presented in Table \ref{tbl:table1}. The images taken with five different exposures for each filter or polarizer were stacked to obtain a single image with a high dynamic range and less noise.

As mentioned above, we carried out observations at multiple sites. One of the objectives of multi-site observations is to detect the evolution of the corona during several tens of minutes, which cannot be detected at a single site owing to the short duration of totality \citep[see][for successful examples during the 2017 eclipse]{2018ApJ...860..142H, 2019ASPC..516..337P, 2020ApJ...888..100B}. In fact, time differences between the two sites were negligible in both eclipses, as observed in Table \ref{tbl:table1}. However, the obtained data are appropriate in verifying the consistency between the results obtained from different sites.

\subsection{Data Processing and Polarization Derivation} \label{subsec:dataproc}

\begin{figure} 
\centerline{\includegraphics[width=1.\textwidth]{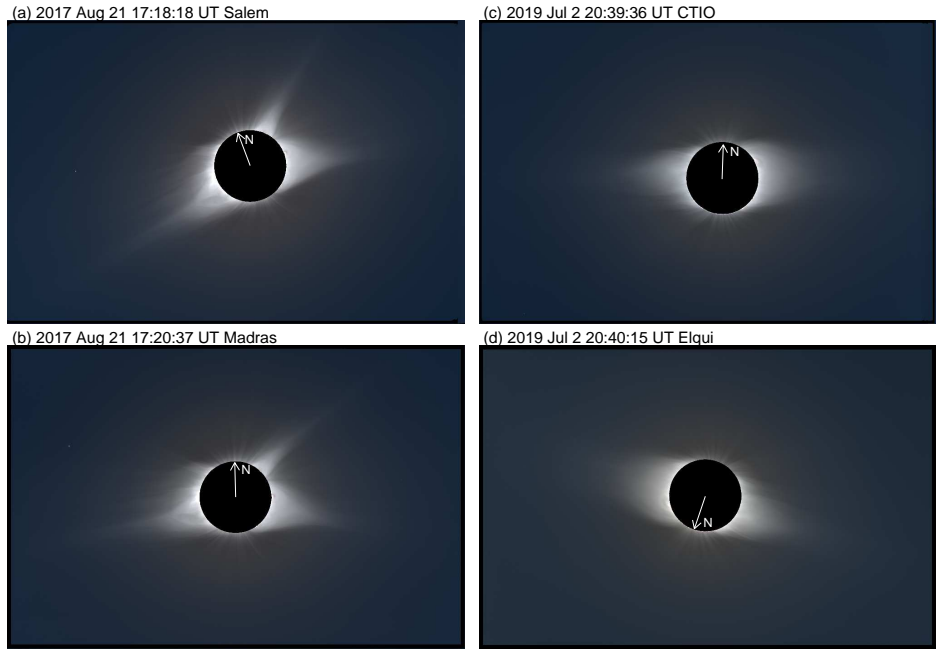}}
\caption{White-light images used for the study obtained during the 2017 and 2019 eclipses. These are Stokes $I$ maps synthesized from signals of three polarization states. The significant radial gradient of the brightness was reduced to expose the coronal structures. The roll angle is not compensated, and the direction of the solar north is depicted with arrows.}\label{fig:fig1}
\end{figure}

To obtain polarization maps, first, we stacked images of each filter/polarizer taken with five different exposures during 3--4 polarization modulations. To successfully stack the images, it is necessary to apply some corrections to the images, in addition to dark and flat corrections.

Digital data recorded by the camera, Canon EOS6D, indicated a slightly non-linear response to the input light intensity (the deviation from the linear response is up to 6 \%), which we corrected. In addition, systematic errors of the exposure times were also corrected. The nominal exposure times were 1, 1/4, 1/16, ... s; however, actual ratios between data numbers in the images taken with adjacent exposure times possibly deviated from 1:4 owing to the systematic errors in exposure times. We estimated the true exposure times based on the ratio of the data numbers in the images  (the deviation from the nominal values were at most 1 \%), and the data numbers were normalized to the values per second using these true exposure times. The non-linear response and the error in the exposure time are closely related to each other, and therefore, we applied these corrections together so that the images taken with different exposure times become consistent to each other.

The position of the Sun in the images taken during totality swayed owing to the tracking errors. We derived relative displacement among the images via the correlation of the coronal fine structures. By applying the correction for this image displacement, we stacked images of each of the filter and polarizers. The overexposed and underexposed parts in the images were omitted for the stacking, and only the data within 4--97 \% of the full pixel-response range were used.

Next, we derived linear polarization signals from the stacked images of the three polarizers. The derivation of the polarization signals as a form of Stokes parameters is explained in, for example, \cite{2021SoPh..296...72I}. 

Raw polarization signals converted from the intensities of three polarization states may contain systematic errors. A reason for these errors could be attributed to the variation in the observation target during the polarization modulation. The polarization modulation with a filter wheel is sequential and relatively slow, and polarization modulated images are taken at different time moments. In such a case, spurious polarization signals can be produced. The coronal brightness was verified to exhibit a negligible change during totality in all the observations, and this type of error can be ignored. However, the brightness of the background sky slightly changed during the totality period; even so, the polarization of the sky is eliminated in our analysis, as described in Section \ref{subsec:removalsky}. 

Nevertheless, we observed systematic errors; the raw linear polarization signals of the corona, which were expected to be tangential to the Sun, exhibited a few kinds of systematic deviations from the tangential direction. The polarization of the sky modifies the orientation of the polarization signals, but the contribution of the sky can be ignored at a small elongation from the solar disk. However, a deviation was found even at a small elongation. This suggests that there were some sort of errors in the data used for the calculation of the polarization signals. Therefore, we applied three types of corrections explained below to minimize the deviation from the tangential direction at 1.1 $R_\odot$, where the coronal brightness is more than a hundred times larger than that of the sky.

The first correction is for the throughput difference between the polarizers. In the 2019 observations, we used a light diffuser to obtain flat-field images, and unpolarized light input from the diffuser was effective in correcting the throughput difference. However, in 2017, we obtained flat-field data with the blue sky. To eliminate the influence of the polarization and brightness gradient of the sky, we took images of the sky by rotating the entire optics (telescope, filter wheel, and camera) every $90^\circ$ around the telescope optical axis. However, the throughput difference could not be eliminated completely. Such a throughput error triggers a systematic deviation in the orientation of the polarization signals. We estimated the throughput difference by assuming that the correct throughputs provide the minimum deviation of the polarization signals from the tangential direction. For the 2017 data, we applied up to 1 \% corrections for the throughputs of the polarizers. Regarding the 2019 data, we confirmed that a minimum deviation is realized without throughput corrections.

The second correction is the positional displacement among the images taken with different polarizers. The images of each polarizer, which were stacked, exhibit almost the same coronal structure; therefore, they can be aligned via the correlation between the coronal fine structures. In contrast, the alignment between the images of different polarization states has an ambiguity, because the images exhibit different brightness distributions owing to the polarization. Therefore, we shifted the polarization images individually to minimize the deviation of the polarization signals from the tangential orientation. The images were shifted approximately $5''$ at the maximum.

The third correction is the orientation error of the filter wheel. Although the relative angle differences between the transmission axes of the polarizers are fixed to be $60^\circ$, the rotation angle of the wheel may deviate from the camera frame (the reference axis) owing to the tolerance of assembling the instruments at the observation sites. Again, we estimated the roll angle of the wheel by assuming that the correct roll angle provides the minimum deviation of the polarization signals from the tangential orientation, and this angle was $2^\circ .5$ at the maximum. 

After applying these three corrections, we obtained Stokes $I$, $Q$, and $U$ data for the R, G, and B channels, respectively, which contain information on the polarization of the corona, as well as the sky background. The obtained Stokes $I$ maps of the corona are presented in Figure \ref{fig:fig1} in the form of the composite of the RGB images. In the later quantitative analysis, we used the data averaged over  $11''.8\times 11''.8$ (See Section \ref{subsec:bpb}); the root-mean-square (RMS) random noise level in the these ``macropixels''  in areas of the lowest brightness of these Stokes $I$ images was approximately 1.5 \%.

To analyze the coronal data quantitatively, the brightness of the corona is required to be expressed in the unit of the solar disk's average brightness, $B_\odot$. We took several images of the solar disk for the reference before the first contact, after the fourth contact (only for 2017) and during the partial eclipse phase, with the same instruments used for coronal observations, but with neutral density filters. We took solar disk images using a stack of two neutral density filters with a total density of approximately 5.6 (transmittance of $10^{-5.6}$). Although it is difficult to measure the spectral transmittance of the filters together (too dense for a measuring instrument), that of each of the filters can be measured with high accuracy. Because the transmittance of the neutral density filters depends on the wavelength, we calculated the effective transmittance for each of the R, G, and B channels using the spectral characteristics of the sensitivity of each color channel of the cameras and that of the solar light \citep[see][]{2012SoPh..279...75H}. The brightness values of the disk derived from these images were compensated for these effective transmittances. Although the transmittance of the neutral density filters and the spectral distribution of the solar light were well defined, the spectral response of the RGB channels of the cameras is expected to include a certain amount of error. This error resulted in the uncertainty of the effective transmittance of 2--20 \% depending on the channels.

The disk-brightness data before and after the totality were interpolated to estimate the brightness at the moment of totality, which was used to calibrate the brightness of the corona. However, there is an uncertainty because of the possible change of the sky condition around the totality. \cite{2012SoPh..279...75H} estimated the uncertainty was $\pm 5$ \% for the 2008 and 2009 eclipses. We should expect that there is the same level of uncertainty in the brightness calibration here.

The brightness calibration was carried out for data obtained at one of the observing sites for each eclipse, where we took disk images with the above-mentioned filters. The brightness scale of the data taken at the other sites were adjusted to the calibrated data. The error of the effective transmittance of the neutral density filters and the possible change of the sky condition around the totality were the largest sources of the uncertainty in the brightness calibration in our analysis.

\begin{figure} 
\centerline{\includegraphics[width=1.\textwidth]{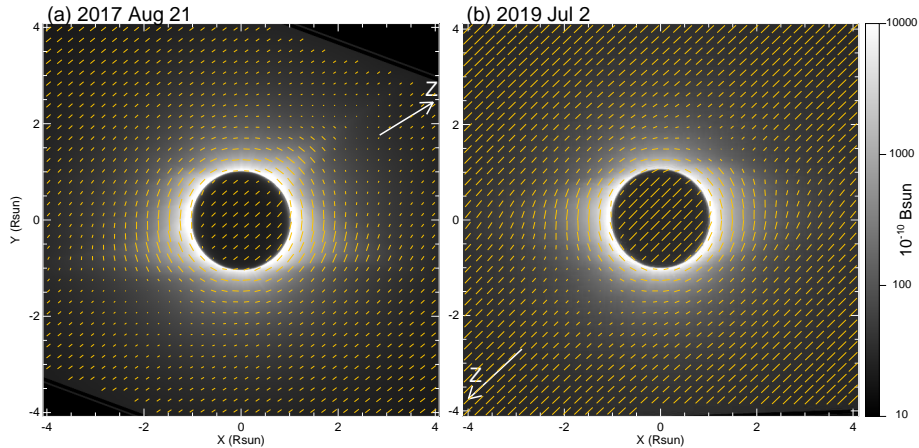}}
\caption{Polarization maps of the G-channel before the removal of the sky background covering $4.1 \times 4.1$ $R_\odot$ area obtained at (a) Salem and (b) CTIO. The grayscale images present Stokes $I$ signals, and the degree and orientation of the linear polarization signals are depicted with orange ticks. The solar north is to the top, and the direction of the zenith is represented with arrows.}\label{fig:fig2}
\end{figure}

Finally, we obtained calibrated polarization maps of the RGB channels taken during the two eclipses observed at the two sites. Representative polarization maps obtained at Salem (for the 2017 eclipse) and at CTIO (for the 2019 eclipse) are presented in Figure \ref{fig:fig2}. In addition to the tangential linear polarization around the solar disk, we can find a background polarization component originating from the sky. In Figure \ref{fig:fig2}, the direction of the zenith is indicated. The polarization of the sky occurs approximately along the vertical direction. It is well-known that the orientation of the sky's polarization in the vicinity of the Sun during totality is almost vertical (e.g., \citealp{1961ApJ...133..616N}; for actual measurements of the sky's polarization during the 2017 eclipse, see \citealp{2019SPIE11132E..0CS}, \citealp{2020ApOpt..59F..41E}, \citealp{2020ApOpt..59F..71S}, and \citealp{2020PASP..132b4202V}). In our observations, the polarization of the sky during the 2019 eclipse is more significant than that of the 2017 one; as observed in Table 1, the altitude of the Sun during the 2019 eclipse was much lower than that during the 2017 eclipse.

\subsection{Removal of the Sky Background} \label{subsec:removalsky}

As seen in Figure \ref{fig:fig2}, the polarization of the sky extends over the corona. To obtain the brightness and polarization of the corona, it is necessary to eliminate sky background, particularly in the part distant from the solar disk. We estimated the polarization and brightness of the sky separately to remove the sky background. Actually, stray light from the instrument and atmosphere also overlaps with the image of the corona; however, it is a minor component of the eclipse observations. For this first-order analysis, we assumed that the stray light is negligibly small.

The linearly polarized component of the sky can be estimated based on the fact that it has a completely different orientation distribution from that of the corona. In Figure \ref{fig:fig2}, the sky appears to have constant polarization; in fact, small variations in the field of view can also be observed. Considering the variation in the field of view, we estimated the polarization of the sky based on two assumptions. The first one is that the Stokes $Q$ and $U$ components of the sky polarization vary linearly in the field of view (in other words, the Stokes $Q$ and $U$ are depicted by a tilted plane). The second assumption is that the orientation of the linear polarization of the corona after the elimination of the sky background is tangential to the Sun. We estimated the $Q$ and $U$ components of the sky, such that the polarization of the K+F corona at 4 $R_\odot$ remaining after the removal of the sky background becomes tangential. As illustrated in Figure \ref{fig:fig2}, up to approximately 4 $R_\odot$, the entire circumference is within the field of view, and at this elongation, the sky is the dominant component in the polarization brightness. 

Next, we estimated the brightness of the sky background based on two other assumptions. The first assumption is that the sky's degree of polarization is constant throughout the field of view, and it is the same value in all of the RGB channels. Formerly derived spectral dependence of polarization on the mid-eclipse sky relatively diverges, and no definite tendency can be determined \citep[refer to the summary by][]{2020ApOpt..59F..41E}. Therefore, we assumed a constant degree of polarization regardless of the RGB channels. The second assumption is that the spectral distribution of the brightness of the K+F corona can be approximated by that of the F-corona at far from the solar disk, because the F-corona becomes dominant beyond 2--3 $R_\odot$. We adopted the spectral distribution of the F-corona presented in \cite{2021ApJ...912...44B}, namely $\lambda^{0.91}$, where $\lambda$ represents the wavelength. We determined the average sky brightness at 4 $R_\odot$, such that the coronal brightness of the RGB channels remaining after the elimination the sky background follows this relationship. The atmospheric absorption, which affects shorter wavelengths more significantly, may slightly alter the apparent color of the F-corona; however, it was ignored for simplicity. These two assumptions or constraints cannot be necessarily satisfied simultaneously for three color-channels. Therefore, we adopted a combination of the approximately same degree of polarization for three color-channels, which precisely provides the assumed spectral distribution of the F-corona. The brightness of the sky in the entire field-of-view was calculated from the derived degree of polarization and polarization brightness of the sky.

\begin{figure} 
\centerline{\includegraphics[width=1.\textwidth]{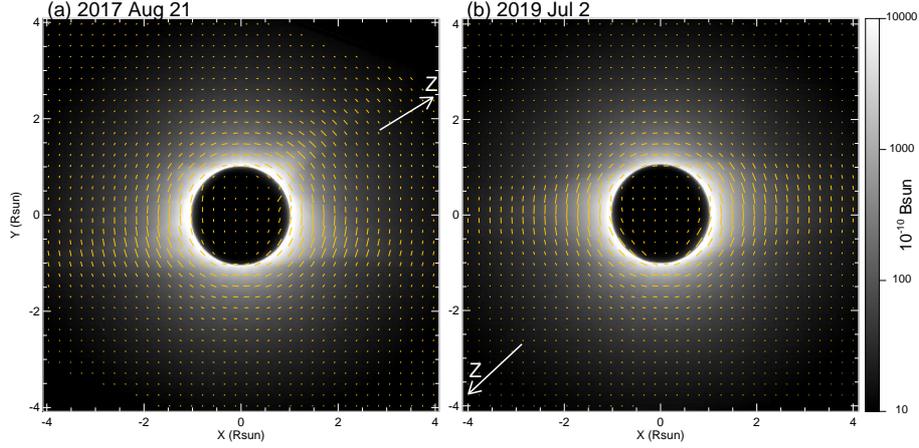}}
\caption{Polarization maps of the K+F corona after the elimination of the sky background, shown in the same manner as in Figure 2.}\label{fig:fig3}
\end{figure}

\begin{figure} 
\centerline{\includegraphics[width=1.\textwidth]{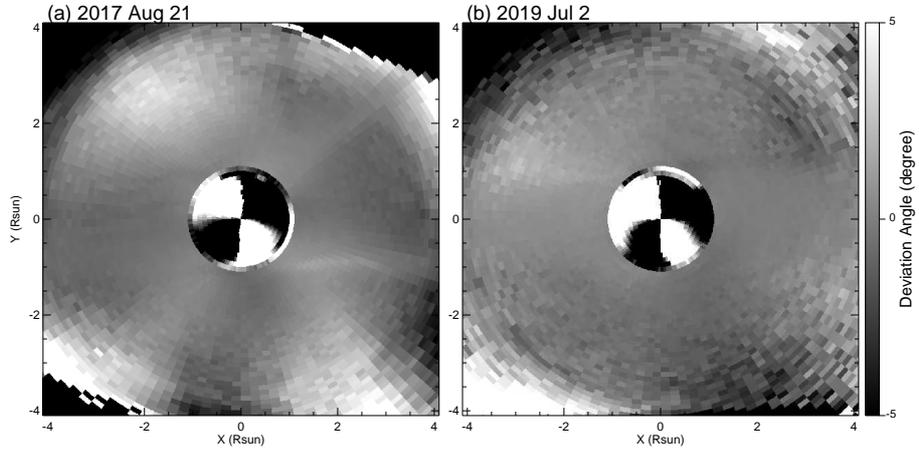}}
\caption{Deviation of the polarization signals' orientation from the tangential direction. To reduce the random noise, the signals are averaged in areas with a position-angle width of $3^\circ$ $\times$ a radial step of 0.1 $R_\odot$.}\label{fig:fig4}
\end{figure}

The estimated sky brightness ranged approximately 8 (R) -- 24 (B) $\times 10^{-10}$ $B_\odot$ during the 2017 eclipse and approximately 14 (R) -- 35 (B) $\times 10^{-10}$ $B_\odot$ during the 2019 eclipse, and the typical RMS fitting error of the sky brightness was 3 \%. In addition, the degrees of polarization were approximately 0.32 and 0.61 during the 2017 and 2019 eclipses, respectively. 

After eliminating the sky background, we obtained the Stokes $I$, $Q$, and $U$ components of the K+F corona; its brightness, polarization brightness, and degree of polarization can be derived from the Stokes components. Figure \ref{fig:fig3} presents the polarization maps of the K+F corona. It can be confirmed that the polarization of the K+F corona is mostly tangential. A deviation remains unvanished from the tangential orientation in the polarization of the corona, as presented in Figure \ref{fig:fig4}. To reduce the random errors owing to shot noise, and to exhibit the large scale pattern, the deviation is averaged 
in areas with a position-angle width of $3^\circ$ $\times$ a radial step of 0.1 $R_\odot$ in Figure \ref{fig:fig4}. Except in the large elongation areas beyond 4 $R_\odot$, the magnitude of the deviation is not much different from the results by \cite{2020PASP..132b4202V}, who derived the polarization of the inner corona during the 2017 eclipse. In the large elongation area, a small error in the estimated polarization of the sky significantly affects the residual polarization. Actually, at 4  $R_\odot$, the RMS error of the polarization brightness after the removal of the sky is as large as 40 \% in the  $11''.8\times 11''.8$ macropixels except in the streamer regions, because the polarized component of the K+F corona comes from only the K-corona, which is much fainter than both the F-corona and the sky at this elongation. Nevertheless, the large scale deviation at 4 $R_\odot$ remains within a few degrees.

\section{Results} \label{sec:results}

Here, we present the brightness ($B_{K+F}$), polarization brightness ($pB$), and degree of polarization ($p_{K+F}$) of the K+F corona derived from our observations of the 2017 and 2019 eclipses. Approximate separation of $B_{K+F}$ into the K- and F-corona components is also presented. In addition, we discuss the comparison between our results and other measurements.

\subsection{Brightness and Polarization Brightness of the K+F Corona} \label{subsec:bpb}

\begin{figure}[t]
\centerline{\includegraphics[width=1\textwidth]{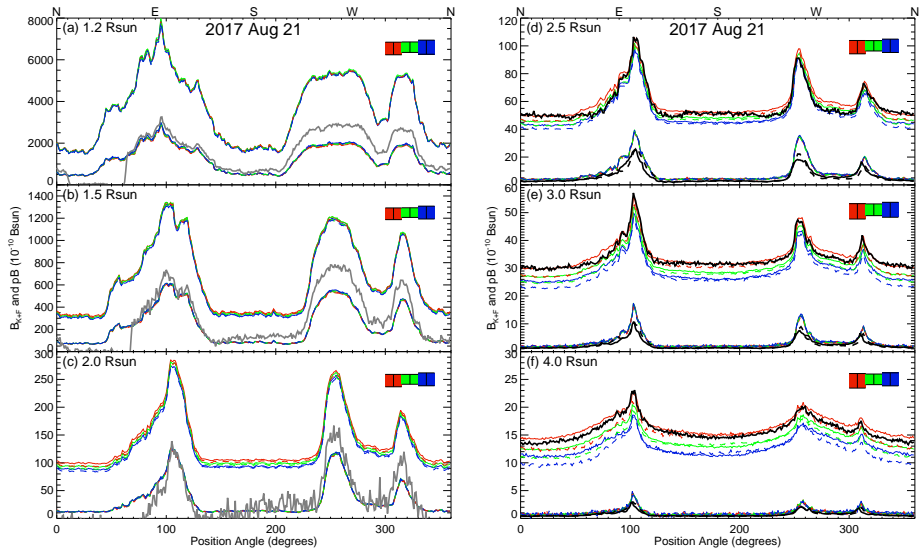}}
\caption{Total brightness of the K+F corona ($B_{K+F}$, upper curves) and its polarization brightness ($pB$, lower curves) at elongations of 1.2--4.0 $R_\odot$ observed during the 2017 eclipse. Red, green, and blue lines correspond to the R, G, and B channels. The data obtained at Salem and Madras are plotted with solid lines and dashed lines, respectively. The typical uncertainty ranges of $B_{K+F}$ of the RGB channels are presented by color bars to the right in each panel. In panels (a)--(c), the $pB$ observations with K-Cor are depicted with solid gray lines. In panels (d)--(f), the $B_{K+F}$ and $pB$ observations with LASCO C2 are presented by black lines (for $pB$, solid lines: 02:54:10, dashed lines: 20:59:55)}\label{fig:fig5}
\end{figure}

\begin{figure} 
\centerline{\includegraphics[width=1\textwidth]{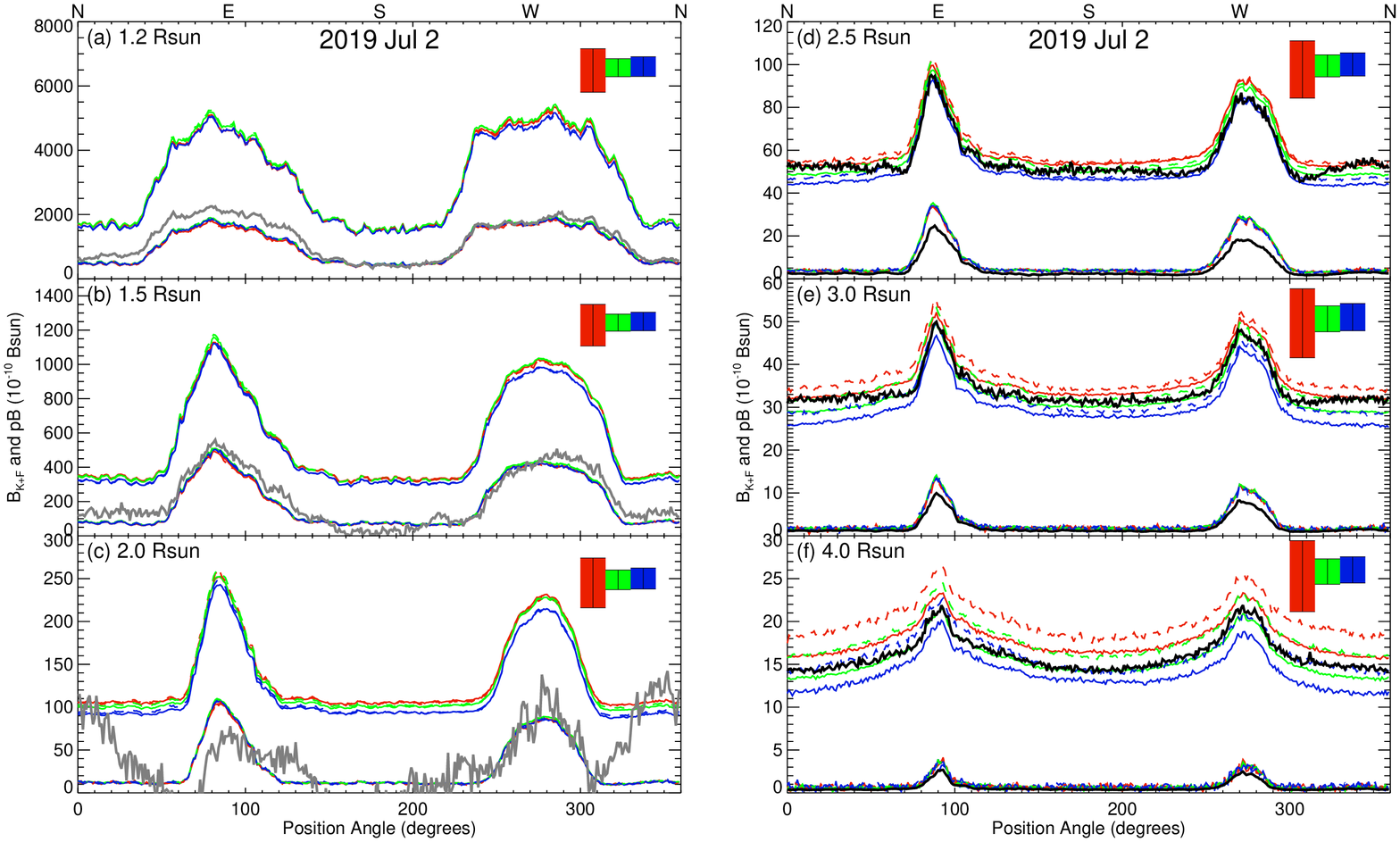}}
\caption{Total brightness of the K+F corona ($B_{K+F}$) and its polarization brightness ($pB$) at elongations of 1.2--4.0 $R_\odot$ observed during the 2019 eclipse plotted in the same manner as in Figure 5. The data obtained at CTIO and Elqui are plotted with solid lines and dashed lines, respectively. The typical uncertainty ranges of $B_{K+F}$ of the RGB channels are presented as well. K-Cor and LASCO C2 data are also plotted in panels (a)--(c) and (d)--(f), respectively, as in Figure 5.}\label{fig:fig6}
\end{figure}

The brightness and polarization brightness of the K+F corona, $B_{K+F}$ and $pB$, at six elongations ranging 1.2--4.0 $R_\odot$, which were derived from the eclipse observations, are presented in Figures \ref{fig:fig5} and \ref{fig:fig6}. The plotted values were averaged over $11''.8\times 11''.8$, which is similar to the pixel size of the LASCO C2 full-resolution data, for comparison. The typical uncertainty ranges of $B_{K+F}$ mentioned in Section \ref{subsec:dataproc} are presented by color bars in Figures \ref{fig:fig5} and \ref{fig:fig6}. Although the three color-channel data were independently calibrated, the results agree well at the low elongations. This fact suggests that the brightness calibration for the three color channels were consistently performed well within the uncertainty ranges. Furthermore, the data of the two sites are basically consistent with each other. The fact that the $pB$ values taken at two sites agree well indicates that the operation of the filter wheels did not experience failure. In addition, the results of the two eclipses, both of which occurred under low solar activity, are similar to each other. With a closer look, it can be determined that the brightness of the 2019 eclipse, which occurred closer to the solar minimum, is relatively smaller than that of the 2017 eclipse. A coronal mass ejection was observed above the east limb during the 2017 eclipse \citep{2020ApJ...888..100B}, and it contributed to the noticeable brightness excess of the corona around the eastern equator. Throughout the height range, the $B_{K+F}$ data are consistent with the brightness of the typical minimum corona previously published by, for example, \cite{1950BAN....11..135V}.

In Figures \ref{fig:fig5} and \ref{fig:fig6}, we overplotted the $pB$ obtained by K-Cor of MLSO (at 17:19:39 for the 2017 data and 20:39:34 for the 2019 data; taken from the MLSO's web page, \url{https://www2.hao.ucar.edu/mlso/mlso-home-page}) for the inner corona up to 2.0 $R_\odot$ in panels (a)--(c). K-Cor measures the polarization brightness of the corona at the height range of 1.05--3.0 $R_\odot$. We binned the K-Cor data with $2\times 2$ pixels, corresponding to $11''.3\times 11''.3$, for comparison. A dip extending around the position angle 10--60$^\circ$ (from the north to the northeast) observed in the K-Cor data of Figure \ref{fig:fig5} occurs because of the Moon. The observation wavelength range of K-Cor is 720--750 nm and it is beyond our observation wavelengths; however, the $pB$ does not significantly change in these wavelength ranges. The eclipse observations and K-Cor data indicate a difference of approximately 30 \% in some parts. However, even if we take the uncertainty of the brightness calibration of the eclipse data into consideration, as a whole, we can find a relatively good agreement at 1.2 and 1.5 $R_\odot$, where the noise owing to the sky background in the K-Cor data is sufficiently small. Structural coincidence between the coronal data obtained at eclipse observations and the K-Cor data was also confirmed by \cite{2019SoPh..294..166J} for the 2017 eclipse and by \cite{2021ApJ...912...44B} for the 2019 eclipse.

In the panels (d)--(f) showing $B_{K+F}$ and $pB$ at elongations of 2.5, 3.0, and 4.0 $R_\odot$ of Figures \ref{fig:fig5} and \ref{fig:fig6}, we plotted LASCO C2 data (obtained from the LASCO-C2 Legacy Archive, \url{http://idoc-lasco.ias.u-psud.fr/sitools/client-portal/doc/}; refer to \citealp{2020SoPh..295...89L}). LASCO C2 takes the $B_{K+F}$ and $pB$ data between 2.2 and 6.5 $R_\odot$. The data at approximately 2.5 $R_\odot$ and farther, which are free from interference fringes owing to the occulting disk, can be adopted for quantitative comparison. The LASCO C2 data used here were obtained with the orange filter (bandpass: 540--640 nm); this wavelength range extends over the G and R channels of our observations. For the 2019 eclipse, the LASCO data obtained at the time close to our eclipse observations are available (20:38:07 for $B_{K+F}$ and 21:13:30 for $pB$). In contrast, for the 2017 eclipse, although the $B_{K+F}$ data close to our observations are available (17:12:07), the closest $pB$ data were obtained at distant times (02:54:10 and 20:59:55). Therefore, for 2017, we plotted both the 02:54:10 and 20:59:55 $pB$ data with solid and dashed lines. The available $pB$ data of LASCO C2 has a pixel size of $23''.8\times 23''.8$ instead of the original pixel size, $11''.9\times 11''.9$, which was used in the $B_{K+F}$ data.

Figures \ref{fig:fig5} and \ref{fig:fig6} demonstrate that the eclipse observations and LASCO C2 data exhibit good agreement for $B_{K+F}$ in both eclipses. However, the $pB$ data of LASCO C2 are systematically smaller than those of the two eclipses, even if we take the error of the brightness calibration into account.

\subsection{Degree of Polarization of the K+F Corona} \label{subsec:degpol}

\begin{figure}[t]
\centerline{\includegraphics[width=1\textwidth]{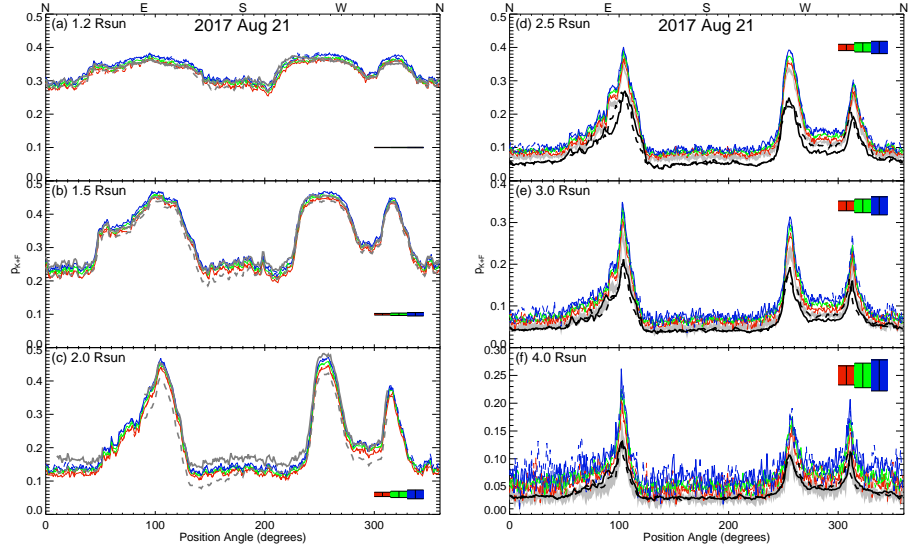}}
\caption{Degree of polarization of the K+F corona, $p_{K+F}$, at elongations of 1.2--4.0 $R_\odot$ observed during the 2017 eclipse. Red, green, and blue lines correspond to the R, G, and B channels. The data obtained at Salem and Madras are plotted with solid lines and dashed lines, respectively. The ranges of the $p_{K+F}$ calculated based on the highest and lowest possible sky brightness are presented by color bars to the right in each panel. In panels (a)--(c), the observations by \cite{2020PASP..132b4202V} are presented with gray lines (solid lines: Bessell B-band, dashed lines: R-band). In panels (d)--(f), the $p_{K+F}$ data derived from LASCO C2 observations are depicted with black lines (solid lines: 02:54:10, dashed lines: 20:59:55). The $p_{K+F}$ data calculated assuming the lowest possible sky brightness are also presented in panels (d)--(f) with gray bands. }\label{fig:fig7}
\end{figure}

\begin{figure} 
\centerline{\includegraphics[width=1\textwidth]{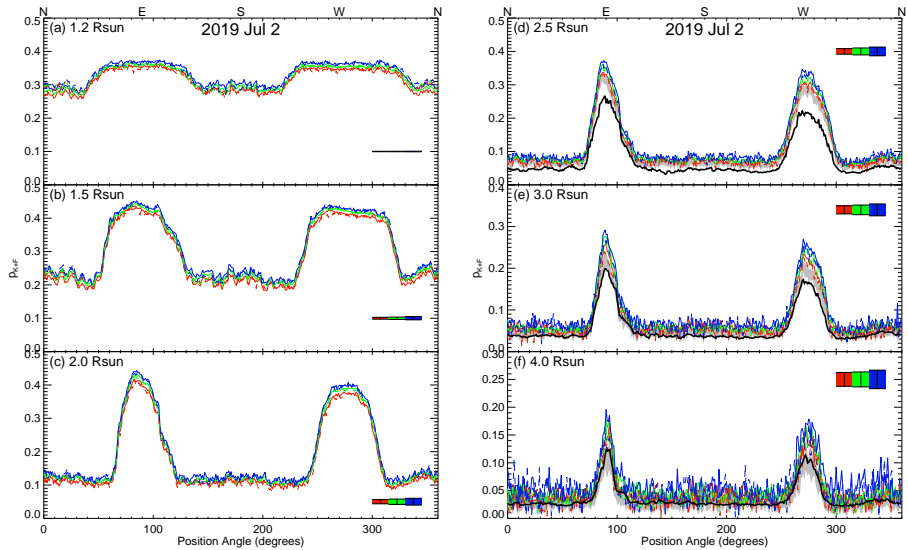}}
\caption{Degree of polarization of the K+F corona $p_{K+F}$ and the RGB channels at elongations of 1.2--4.0 $R_\odot$, observed during the 2019 eclipse. The data at CTIO and Elqui are plotted with solid and dashed lines, respectively. The ranges of the $p_{K+F}$ calculated based on the highest and lowest possible sky brightness are presented by color bars. In panels (d)--(f), the data of LASCO C2 and the eclipse $p_{K+F}$ that are calculated assuming the lowest possible sky brightness are also presented similarly to Figure \ref{fig:fig7}.}\label{fig:fig8}
\end{figure}

Figures \ref{fig:fig7} and \ref{fig:fig8} represent the degree of polarization of the K+F corona, $p_{K+F} = pB/B_{K+F}$. For the 2017 eclipse, the polarization data presented by \cite{2020PASP..132b4202V} are also depicted for elongations up to 2 $R_\odot$, and for both eclipses, the LASCO C2 data are presented for elongations beyond 2 $R_\odot$. For the 2017 eclipse when the LASCO $pB$ data are obtained separately from the eclipse time, the LASCO $pB$ data at 02:54:10 and 20:59:55 were divided by the LASCO $B_{K+F}$ data, taken closely to the $pB$ data, at 02:48:08 and 20:48:07, respectively.

\cite{2020PASP..132b4202V} observed the 2017 eclipse with Rochester Institute of Technology Polarization Imaging Camera (RITPIC), a special polarization imaging camera with pixel-based polarizers on the detector, at Madras, Oregon, USA, where one of our observations was carried out. They took coronal images with approximately $4\times 4$ $R_\odot$ field of view in the Bessell B and R wavelength bands (approximately correspond to the B and R channels of DSLR cameras), and derived the polarization parameters. Their $p_{K+F}$ data were binned into a $11''.5\times 11''.5$ pixel in Figure \ref{fig:fig7}. Their data agree with ours very well. 

However, reflecting the discrepancy observed in the $pB$ data of Figures \ref{fig:fig5} and \ref{fig:fig6}, the $p_{K+F}$ data of LASCO C2 exhibits a systematic difference from the eclipse data. The $p_{K+F}$ values of LASCO C2 are 0.68 times those of the eclipses in average.

Unlike $B_{K+F}$ and $pB$, which need to be calibrated using the average brightness of the solar disk, $p_{K+F}$ can be determined free from the brightness calibration, because its derivation can be done based on the relative variation of the brightness on images taken with linear polarizers with different polarization orientations. Furthermore, linear polarizers have a high polarimetric performance. Ordinary plastic film linear polarizers show a high extinction ratio ($10^3$ -- $10^5$) over a wide wavelength range, and their polarization properties are little affected by incident angle of light or temperature. Therefore, the uncertainty of $<$ 2 \% in $p_{K+F}$ derived by \cite{2020PASP..132b4202V} can be considered as a typical error in the measurements using ordinary linear polarizers.

Besides the uncertainty of the brightness calibration, the eclipse $B_{K+F}$ data have an ambiguity triggered by the error of the sky brightness. Here we present the uncertainty range of $p_{K+F}$ determined by taking the extreme error of the sky brightness into account. The highest possible sky brightness is the measured total brightness near the outer edge of the field of view, approximately at 6.5 $R_\odot$; even at this elongation, $B_{K+F}$ is not zero, and therefore, the true sky brightness must be lower than the total brightness at this elongation.
The lowest possible sky brightness is the polarization brightness of the sky; this means that the sky's degree of polarization is 100 \%. 
The ranges of the $p_{K+F}$ between those calculated based on the highest and lowest sky brightness are presented in Figures \ref{fig:fig7} and \ref{fig:fig8} by color bars. 

If the true sky brightness is lower than the estimated one, the eclipse $B_{K+F}$ should be larger, and it leads to a smaller $p_{K+F}$. To verify whether the discrepancy between the eclipse observations and LASCO data was triggered by the ambiguity of the sky's brightness, we presented $p_{K+F}$ calculated assuming the lowest possible sky brightness in the panels (d)--(f) in Figures \ref{fig:fig7} and \ref{fig:fig8} with gray bands. The range between the maximum and minimum of the calculated $p_{K+F}$ is depicted by the bands. In the case of the 2.5 and 3.0 $R_\odot$, the degree of polarization based on the low sky-brightness assumption is still higher than the LASCO values. Finally at 4.0 $R_\odot$, it approaches the LASCO values. Therefore, even an unrealistic assumption that the degree of polarization of the sky is 100 \% cannot explain the discrepancy except at the 4.0 $R_\odot$.

\cite{2020SoPh..295...89L} discussed the consistency between LASCO C2 data and the eclipse data presented by \cite{2020PASP..132b4202V}. 
However, their eclipse data cover up to 2 $R_\odot$, and the LASCO C2 data provide reliable results beyond 2.5 $R_\odot$. There is no overlap, and it appears difficult to reach a definite conclusion. However, our wide-field data suggest that a systematic discrepancy exists between the eclipse data and the LASCO data in terms of the polarization, in both the 2017 and 2019 eclipses.

\subsection{Approximate Estimation of the Brightness and Polarization of the K- and F-Corona} \label{subsec:kandf}

\begin{figure}[t]
\centerline{\includegraphics[width=1\textwidth]{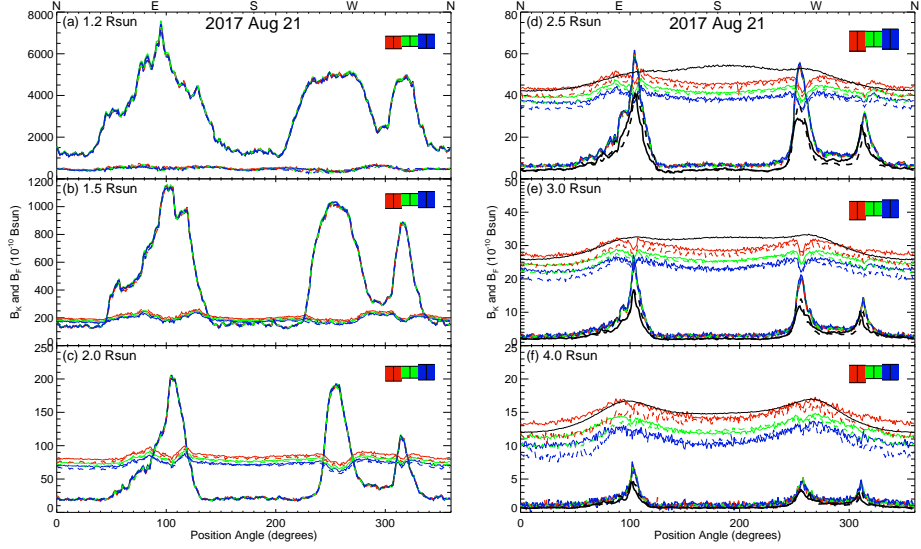}}
\caption{Estimated brightness of the K-corona, $B_K$ (thick color lines), and F-corona, $B_F$ (thin color lines), at elongations of 1.2--4.0 $R_\odot$ for the 2017 eclipse. Red, green, and blue lines correspond to the R, G, and B channels. The data obtained at Salem and Madras are plotted with solid and dashed lines, respectively. The typical uncertainty ranges of $B_{K+F}$ of the RGB channels are presented by color bars to the right in each panel. In panels (d)--(f), the $B_K$ and $B_F$ data of LASCO C2 are represented by thick and thin black lines (for $B_K$, solid lines: 02:54:10, dashed lines: 20:59:55).}\label{fig:fig9}
\end{figure}

\begin{figure} 
\centerline{\includegraphics[width=1\textwidth]{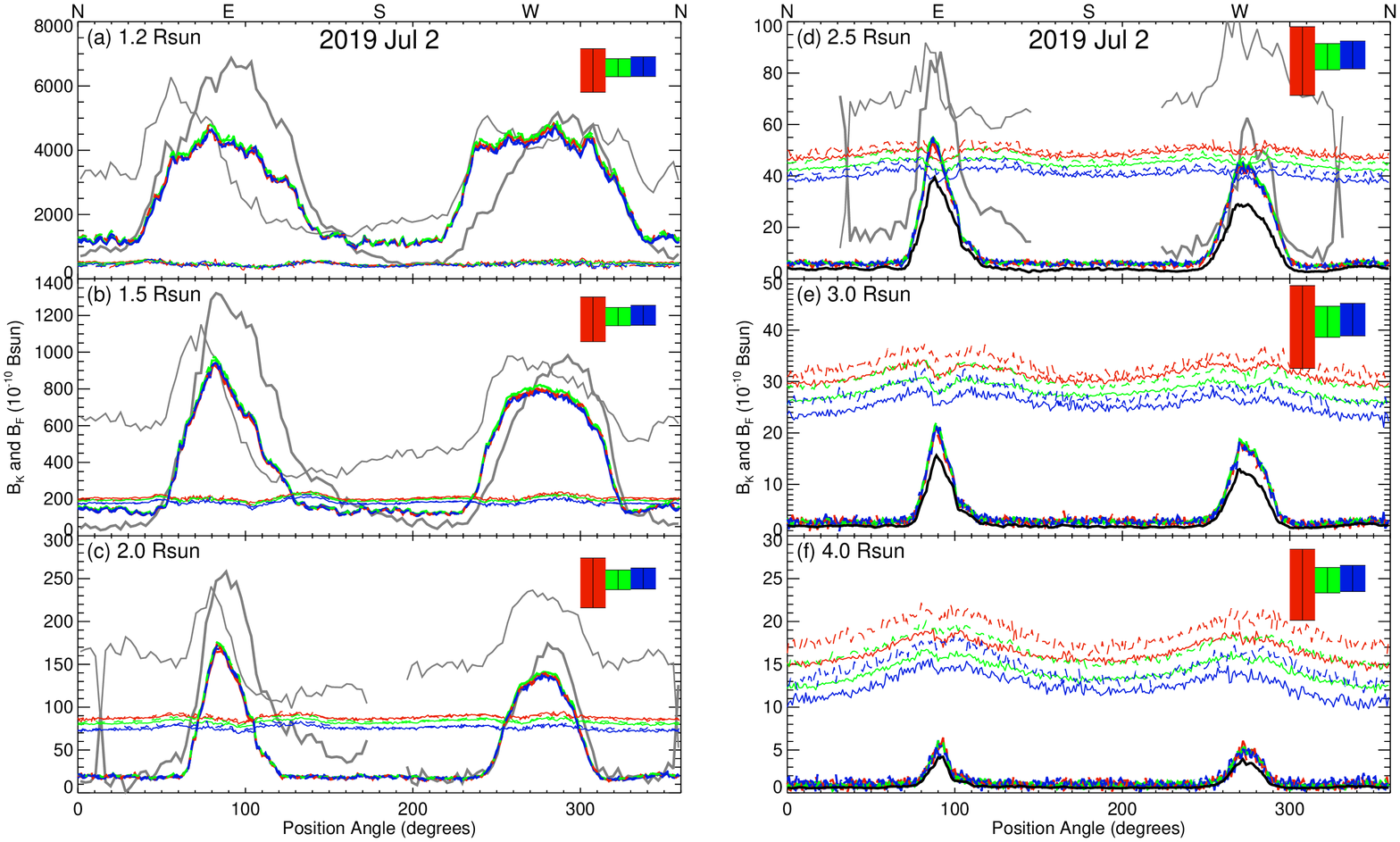}}
\caption{Estimated brightness of the K-corona, $B_K$ (thick color lines), and F-corona, $B_F$ (thin color lines), of the RGB channels at elongations of 1.2--4.0 $R_\odot$ for the 2019 eclipse drawn in the same manner as in Figure \ref{fig:fig9}. The data at CTIO and Elqui are plotted with solid and dashed lines, respectively. The typical uncertainty ranges of $B_{K+F}$ of the RGB channels are presented as well. In panels (a)--(d), the $B_K$ and $B_F$ values by \cite{2021ApJ...912...44B} in unit of the brightness of the solar disk center, are represented by thick and thin gray lines, respectively. Furthermore, in panels (d)--(f), the $B_K$ data of LASCO C2 are represented by thick black lines.}\label{fig:fig10}
\end{figure}

Next, we consider the isolation of the K- and F-corona from the observed K+F corona based on the $B_{K+F}$ and $pB$ data. 

The $B_{K+F}$ is the sum of the brightness of the K-corona, $B_K$, and that of the F-corona, $B_F$, such as, $B_{K+F} = B_K + B_F$. The $pB (= p_{K+F}B_{K+F})$ can be written as $pB = p_KB_K+p_FB_F$ using the degree of polarization of the K- and F-corona, $p_K$ and $p_F$. In the range of elongations discussed here, $p_F$ can be assumed to be zero. Therefore, $pB= p_KB_K$, and $B_{K+F} = pB/p_K+B_F$. To isolate the K- and F-corona, it is necessary to know $p_K$ and $B_F$. 

We carried out the isolation of the K- and F-corona on the assumption that both the $p_K$ and $B_F$ solely depend on the elongation from the disk center. Actually, a calculation of $p_K$ by \cite{2020SoPh..295...89L}, based on the electron density model by \cite{1938AN....267..273B}, indicates that the $p_K$ solely depends on the elongation (refer to their Figure \ref{fig:fig2}). They indicate that the $p_K$ asymptotically reaches approximately 0.64 at large elongations. Furthermore, as demonstrated by \cite{2021ApJ...912...44B}, a realistic electron-density distribution prediction of the corona for the 2019 eclipse (PSI-MHD model; magnetohydrodynamic model by Predictive Science Inc.) provides an approximately constant $p_K$ regardless of the position angle at each elongation near the limb ($<$1.5 $R_\odot$). Beyond this elongation, $p_K$ exhibits fluctuation; there is a tendency that the bright coronal streamers show a high degree of polarization. The PSI's MHD modeling (\url{http://www.predsci.com/corona/coronal\_modeling.html}) provides the $B_K$ and $pB$ distributions for other eclipses as well, and we can confirm that they exhibit a similar tendency to that of the 2019 eclipse.
Regarding the F-corona, although the F-corona is known to extend along the ecliptic, in the inner corona, the ellipticity of the F-corona is negligible (e.g., \citealp{1977SoPh...55..121S}; see also \citealp{1985ASSL..119...63K}). 

If we adopt a constant $p_K$ regardless of the position angle at an elongation $r$, $B_F$ can be expressed using the observed values of $B_{K+F}$ and $pB$ as 
$$B_F(\theta, r)= B_{K+F} (\theta, r)-pB(\theta, r)/ p_K(r)$$
where $\theta$ represents the position angle. The assumption mentioned above requires $B_F(\theta, r)$ to be independent of $\theta$. Therefore, we determined $p_K(r)$, such that the fluctuation (standard deviation) of $B_F(\theta, r)$ calculated by the above equation becomes minimum. The isolated $B_K = pB/ p_K$ and $B_F$ are presented in Figures \ref{fig:fig9} and \ref{fig:fig10}. In panels (a)--(c), $B_K $ and $B_F$ derived independently for the RGB channels are presented for the elongations at 1.2, 1.5, and 2.0 $R_\odot$. For larger elongations, we calculated $B_K$ and $B_F$ on the assumption that $p_K=0.64$ \citep{2020SoPh..295...89L} for all the color channels, and they are depicted in panels (d)--(f). The uncertainty ranges of the $B_{K+F}$ presented in Figures \ref{fig:fig5} and \ref{fig:fig6} are displayed again for the comparison.

The color of the F-corona  presumed from the $B_F$ curves of the three color channels in Figures \ref{fig:fig9} and \ref{fig:fig10} appears approximately uniform at each elongation; the rather rough assumptions explained above successfully reproduced the fixed color of the F-corona. Looking at the details, we can find that 
the $B_F$ data from the eclipses at 2.5 $R_\odot$ or farther exhibit two broad increases around the equator (position angle P.A.$=90^\circ$ and $270^\circ$); these are caused by the elliptical distribution of the F-corona at large elongations. In addition, dips in the eclipse $B_F$ data can be found at the peaks of the K-corona streamers. This fact probably indicates that the true degree of polarization around the peaks of streamers is relatively higher than the surroundings, as presented by the PSI-MHD model \citep{2021ApJ...912...44B}.

There are some other measurements of $B_K$ and $B_F$. The data from the LASCO C2 catalog are presented in Figures \ref{fig:fig9} and \ref{fig:fig10} (panels (d)--(f) in both the figures) with thick and thin black lines.  The LASCO $B_F$ data for the 2017 presents fairly good coincidence to our results ($B_F$ data for the 2019 eclipse are unavailable). The LASCO $B_K$ data were estimated based on the polarization. They basically show similar behavior to our results, except that they are systematically lower than our results. 

In Figure \ref{fig:fig10}(a)--(d), we presented $B_K $ and $B_F$ estimated by \citeauthor{2021ApJ...912...44B} (\citeyear{2021ApJ...912...44B}; numerical data are from \url{https://www.ifa.hawaii.edu/SolarEclipseData/NarrowCont.html}) for the 2019 eclipse with thick and thin gray lines. Their $B_K $ and $B_F$, binned with a position-angle width of $3^\circ$ and a radial step of 0.02 $R_\odot$, are expressed in unit of the brightness of the solar disk center, and therefore, those in unit of the average brightness of the disk actually about 20 \% higher than the presented values. They obtained narrow-band coronal images at several wavelengths, and calibrated them using the MLSO K-Cor measurements of $pB$ and the relation between $pB$ and $B_K$ derived from the PSI-MHD model. Furthermore, they estimated $B_F$ based on the spectral characteristics of the narrow-band coronal images. Even if we take the uncertainty of the data into account, their $B_K$ data are larger than our results and the LASCO data, while their $B_K$ data show a similar position-angle dependence to that of the other observations. The F-corona estimated by \cite{2021ApJ...912...44B} are not only significantly higher than our estimation, but also shows a notable dependence on the position angle. In addition, \cite{2021ApJ...912...44B} suggested that the F-corona is polarized. The polarization of the F-corona was also pointed out by \cite{2020ApJ...893...57M}; they estimated the degree of polarization of $6.6\pm0.9$ \% between 4 and 5.5 $R_\odot$. This means that $pB$ at 4 $R_\odot$ shown in Figures \ref{fig:fig5} and \ref{fig:fig6} is mostly coming from the F-corona, and $B_K$ at 4 $R_\odot$ is much smaller than that presented in Figures \ref{fig:fig9} and \ref{fig:fig10}. Polarization of the F-corona potentially gives significant impact on the estimations of the K-corona brightness particularly at the large elongations.

\begin{figure}[t] 
\centerline{\includegraphics[width=1\textwidth]{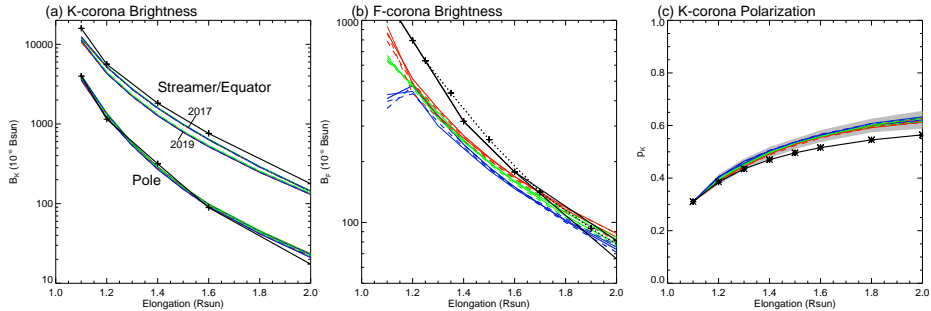}}
\caption{Estimated brightness of the K- and F-corona, $B_K$ and $B_F$, and degree of polarization of the K-corona, $p_K$. Red, green, and blue lines correspond to the R, G, and B channels. Thick and thin color lines represent the 2017 and 2019 data. The data at Salem and CTIO are plotted with solid color lines, and those at Madras and Elqui are plotted with dashed color lines. In panel (a), $B_K$ around the streamers and poles, averaged in sectors with the width of 30$^\circ$, are shown. The equator and pole $B_K$ values by \cite{1985ASSL..119...63K} are also depicted by plus symbols connected by solid lines. In panel (b), the $B_F$ values averaged over all position angles are shown. In addition, the $B_F$ values by \cite{1985ASSL..119...63K} and \cite{1982A&A...112..241D} are depicted by plus symbols connected by solid lines (equator and pole) and a dotted line, respectively. In panel (c), the estimated $p_K$ is shown with color lines, and in addition, the gray band behind the lines for the $p_K$ indicates the one-sigma ambiguity range inferred from the standard deviation of $B_F$. The $p_K$ values predicted by the PSI-MHD models \citep{2021ApJ...912...44B} are represented by star symbols connected by a solid line.}\label{fig:fig11}
\end{figure}

In Figure \ref{fig:fig11}, the estimated $B_K$ averaged around the streamers and poles, $B_F$ averaged over all position angles, and $p_K$ at several elongations up to 2 $R_\odot$ are presented. The data for the two sites, two eclipses and three color-channels are plotted together; they mostly overlap each other except the $B_K$ at streamers, and this indicates the consistency among the independently derived values. The systematic difference at the streamers in 2017 and 2019 is approximately 20 \%. As mentioned in Section \ref{subsec:bpb}, this corresponds to the higher activity level of the corona during the 2017 eclipse than that during the 2019 eclipse. In spite of this difference, the $B_K$ values are basically similar to those for the equator and poles by \citeauthor{1985ASSL..119...63K} (\citeyear{1985ASSL..119...63K}; presented with plus symbols in panel (a)). 

The $B_F$ data in Figure \ref{fig:fig11}(b) are also similar to the previous measurements by \cite{1985ASSL..119...63K} and \cite{1982A&A...112..241D} (shown with plus symbols in panel (b)) except at 1.1 $R_\odot$. The fitting at 1.1 $R_\odot$ failed because the brightness of the F-corona was far smaller than that of the K-corona.

In the $B_F$ data of the three color-channels shown in Figure \ref{fig:fig11}(b), the brightest was the R channel and followed in order by G and B. It is natural because the F-corona is red; however, it is not as red as expected from the relation $\lambda^{0.91}$ \citep{2021ApJ...912...44B}, which we adopted in Section \ref{subsec:removalsky} for the spectral distribution of the F-corona at 4 $R_\odot$. On the contrary, the RGB brightness of the K+F corona beyond 4 $R_\odot$ (virtually pure F-corona) inferred from the sky brightness derived in Section \ref{subsec:removalsky} is redder than that expected from the relation $\lambda^{0.91}$. These facts suggest the possibility that the assumption of the spectral distribution of the F-corona is too red (namely the eliminated sky is too blue). Actually, adopting the relation $\lambda^{0.48}$ for the F-corona and applying it to estimate the sky brightness at 4 $R_\odot$, we can obtain more consistent spectral distribution of the F-corona from 2 $R_\odot$ to beyond 4 $R_\odot$ (however, within 2 $R_\odot$, an even smaller exponent is required). However, the estimation of the F-corona brightness is based on various assumptions as mentioned above, and the results from the two sites of the two eclipses show considerable dispersion regarding the color of the F-corona; therefore, further investigations are needed to judge whether the spectral distribution of the F-corona should be updated or not.

The $p_K$ values presented in Figure \ref{fig:fig11}(c) are not significantly different from the average of the $p_K$ predicted by the PSI-MHD model \citep[][presented with stars in panel (c)]{2021ApJ...912...44B}. They are also similar to the $p_K$ derived from a simple model calculation by \cite{2020SoPh..295...89L}. 

Consequently, we can conclude that the plausible $B_K$, $B_F$, and $p_K$ values were successfully derived from the observed $B_{F+K}$ and $pB$ self-consistently without external calibration data, only assuming that the $p_K$ and $B_F$ solely depend on the elongation from the disk center. The $p_K$ and $B_F$ derived here should be considered a type of average, because the assumption for the $p_K$ and $B_F$ is relatively rough. If the $B_F$ has significant dependence on the position angle or the F-corona is substantially polarized as pointed by \cite{2021ApJ...912...44B} and \cite{2020ApJ...893...57M}, the results need to be reexamined.
Nevertheless, the results support the reliability of the $B_{F+K}$ and $pB$ measured in the two eclipses. Generally, to verify whether the measured $B_{K+F}$ and $pB$ values are correlated with the plausible $B_F$, $B_K$, and $p_K$ values will be a suitable test for the consistency between $B_{K+F}$ and $pB$ values.

\section{Summary and Discussion} \label{sec:summary}

We carried out observations of the total solar eclipses in 2017 and 2019 including polarization measurements at two sites for each eclipse. The $B_{K+F}$ and $pB$ data from just above the limb to approximately 4 $R_\odot$ in three color-channels were obtained under a low sky background. Furthermore, we approximately isolated the brightness of the K- and F-corona. These are basic parameters for the quantitative analyses of the corona such as the derivation the correct electron density distribution and the monitoring of the long-term coronal activity variation. The results from the two sites coincide well, and the two eclipses, which occurred under the low solar activity, provided consistent results. These facts increase the reliability of our measurements.

However, the comparison of the measurements with other observations gave considerably scattered results.
For the inner corona, our results exhibit overall coincidence with the $pB$ measured by K-Cor. The $B_K$ estimated by \cite{2021ApJ...912...44B} and our results show similar brightness distributions, though the data by \cite{2021ApJ...912...44B} are systematically larger than our results. For the $p_{K+F}$, our data show very good coincidence with the those measured by \cite{2020PASP..132b4202V}. 
The estimation of the brightness of the F-corona in the inner corona again shows significant scatter. While our results up to 2 $R_\odot$ are not much different from the previous measurements by \cite{1985ASSL..119...63K} and \cite{1982A&A...112..241D}, \cite{2021ApJ...912...44B} obtained much larger brightness and strong dependence to the position angle of the F-corona. Furthermore, \cite{2021ApJ...912...44B} and \cite{2020ApJ...893...57M} suggest that the F-corona is polarized. Because the separation of the K- and F-corona have been done on the assumption that the F-corona is unpolarized, the polarization of the F-corona, if any, affects the estimation of the brightness of the K-corona.

For the outer corona, although the $B_{K+F}$ of our measurements and those of LASCO C2 exhibit good coincidence, the $pB$, $p_{K+F}$, and $B_K$ exhibit systematic discrepancy, even if the uncertainty is taken into consideration. 
\cite{2020SoPh..295...89L} compared the LASCO C2 polarization data with data obtained during several previous eclipses. The data from the 1998 and 1999 eclipses are consistent with the $p_{K+F}$ measured with LASCO C2. In contrast, the data taken during the 2006 March 29 eclipse exhibits the tendency that the degree of polarization measured with LASCO C2 is smaller than that of the eclipse data; they stated that the LASCO $p_{K+F}$ was 0.60--0.64 of the eclipse observations. This is comparable to our case, where the factor of 0.68 was derived. This fact supports the conclusion that a systematic discrepancy exists between the $pB$, $p_{K+F}$, and $B_K$ derived from the two eclipses and those of LASCO C2.

In such a situation, to quantitatively obtain accurate coronal properties, it is important to establish the consistent calibration for the brightness and polarization of the corona in various observational methods. Although the solar eclipses can be observed at only very limited chances, once in a couple of years, they provide a unique opportunity to obtain irreplaceable data. 

As listed below, our observation method for the total solar eclipses has some features that are suitable to calibrate and separate the coronal components self-consistently and to do the intercomparison among various kinds of data, such as eclipse observations, observations with spaceborne and ground-based coronagraphs, and MHD simulations.
\begin{itemize}
\item The field of view $>$ 4 $R_\odot$ enabled us to compare the results with other observations of both the inner and outer corona, such as by K-Cor and LASCO C2. 

\item The brightness calibration was done using the solar disk images taken with the same instruments used for coronal observations, independently from other measurements.

\item Measurements of both the brightness and polarization of the corona are helpful to remove the background sky, and to separate the K- and F-corona, self-consistently.

\item The data of three color channels are also helpful to separate the K-corona, F-corona, and sky, which have different spectral characteristics. The capability of the simultaneous imaging in the three color-channels became possible by the employment of commercial DSLR cameras.

\item The observations carried out at multiple sites greatly contribute to the confirmation of the consistency between the results.

\item We have attempted to obtain the data using similar instruments at some eclipses, and succeeded to observe the corona at two eclipses. It enabled us to check the consistency of the results and the differences of the corona in the two eclipses. 

\end{itemize}

The multi-site polarimetry is a particularly notable feature in our observations. It was important that we used a traditional filter wheels for the polarization modulation. Some specialized instruments, which can obtain linear polarization information instantaneously, have been introduced for the eclipse observations. The PolarCam and RITPIC adopted by \cite{2019SoPh..294..166J} and \cite{2020PASP..132b4202V}, respectively, have pixel-based polarizers on their detectors. The Leiden Eclipse Imaging Polarimeter used by \cite{2020ApOpt..59F..71S} has three DSLR cameras equipped with linear polarizers set to have three different transmission axis orientations. However, it will be challenging to dispatch expeditions to multiple sites with such instruments. Conversely, it is relatively easy to setup several instruments, including a simple filter wheel and DSLR camera. In addition, these instruments can be transported and operated by a single person. As a result, we successfully took data at two sites in collaboration with amateur observers.

Eclipse data covering a wide field of view with a low noise level are especially important to connect various coronal observations and eventually obtain the correctly calibrated coronal brightness and polarization data. Therefore, the observations of the total solar eclipses still increases its value.

%
\begin{acks}
We thank all the observers that participated in our multi-site observation programs for the 2017 and 2019 total solar eclipses.
The Association of Universities for Research in Astronomy accepted our proposal to observe the 2019 eclipse at CTIO, and the observation was carried out with the help of the CTIO staff and Dr. S. Morita of NAOJ, a member of the observing team.
We are grateful to Dr. D. Vorobiev for providing the polarization data of the 2017 eclipse. The numerical data of the 2019 eclipse presented by \cite{2021ApJ...912...44B} are thankfully available in the total solar eclipse database of the University of Hawaii's Institute for Astronomy.
The K-Cor data were courtesy of the Mauna Loa Solar Observatory, operated by the High Altitude Observatory, as part of the National Center for Atmospheric Research (NCAR). NCAR is supported by the National Science Foundation. 
This work makes use of the LASCO-C2 legacy archive data produced by the LASCO-C2 team at the Laboratoire d'Astrophysique de Marseille and the Laboratoire Atmosph\`eres, Milieux, Observations Spatiales, both funded by the Centre National d'Etudes Spatiales (CNES). LASCO was built by a consortium of the Naval Research Laboratory, USA, the Laboratoire d'Astrophysique de Marseille (formerly Laboratoire d'Astronomie Spatiale), France, the Max-Planck-Institut f\"ur Sonnensystemforschung (formerly Max Planck Institute f\"ur Aeronomie), Germany, and the School of Physics and Astronomy, University of Birmingham, UK. SOHO is a project of international cooperation between ESA and NASA. The authors are grateful to the anonymous referee for his/her careful reading and helpful comments.

\end{acks}

\vspace\baselineskip
\noindent{\bf Disclosure of Potential Conflicts of Interest} The authors declare that they have no conflicts of interest.

%
%
%

\end{article} 
\end{document}